\documentclass{article}

\usepackage{arxiv}

\usepackage[utf8]{inputenc} % allow utf-8 input
\usepackage[T1]{fontenc}    % use 8-bit T1 fonts
\usepackage{hyperref}       % hyperlinks
\usepackage{url}            % simple URL typesetting
\usepackage{booktabs}       % professional-quality tables
\usepackage{amsmath}        % equation and align environments
\usepackage{amsfonts}       % blackboard math symbols
\usepackage{amssymb}        % extra math symbols
\usepackage{microtype}      % microtypography
\usepackage{graphicx}       % figures
\usepackage[numbers,sort&compress]{natbib} % numeric citations, as in the journal version
\usepackage{doi}            % hyperlinked DOIs (used in the .bib notes)
\usepackage{hyperref}
\usepackage{orcidlink}

\title{Co-Design Optimization for Data Center Cooling System via Digital Twin}

\renewcommand{\shorttitle}{Co-Design Optimization for Data Center Cooling via Digital Twin}

\author{Shrenik Jadhav \orcidlink{0009-0003-6906-7465} \\
	Department of Computer and Information Science\\
	University of Michigan-Dearborn\\
	4901 Evergreen Rd, Dearborn, MI, USA \\
	\And
	Zheng Liu\thanks{Corresponding author.} \orcidlink{0000-0003-4869-8893} \\
	Department of Industrial and Manufacturing Systems Engineering\\
	University of Michigan-Dearborn\\
	4901 Evergreen Rd, Dearborn, MI, USA \\
	\texttt{zhengtl@umich.edu} \\
}

\hypersetup{
	pdfauthor={Shrenik Jadhav and Zheng Liu},
	pdftitle={Co-Design Optimization for Data Center Cooling System via Digital Twin},
	pdfkeywords={Data center, thermal management, co-design optimization, digital twin},
	pdfsubject={Co-design optimization of the cooling system for an exascale supercomputer},
}

\begin{document}

\maketitle

\begin{abstract}
Liquid-cooled exascale supercomputers dissipate heat through cooling plants organized as multiple parallel subloops, but how to allocate coolant distribution units (CDUs) across subloops and how to distribute flow among them has not been systematically addressed for facilities at this scale. This paper presents a three-layer optimization framework that jointly determines the integer partition of CDUs across subloops, the continuous flow fraction allocation, and the per-timestep co-design optimization of total flow rate and supply temperature subject to per-subloop thermal safety constraints. The Modelica simulation model is built based on the data of the Frontier exascale supercomputer at Oak Ridge National Laboratory. By developing a reduced-order surrogate model, all 611 feasible partitions of 25 CDUs are evaluated across the full year operational dataset of 49{,}353 timesteps. Three progressively richer operational strategies are compared, ranging from flow control optimization to full three-layer co-design optimization with dynamically adjusted flow fractions. The optimal design within the surrogate optimization problem is a two-subloop plant achieving 35.48\% annual cooling energy savings, only 0.18\% above the current three-subloop design at 35.30\%. Most of the savings are delivered by supervisory co-optimization of total flow rate and supply temperature; the distinct role of flow fraction optimization is design robustness rather than additional raw savings. Flow fraction optimization compensates for any feasible CDU-to-subloop assignment, reducing the design sensitivity by 93\% and providing a low-cost software-only pathway to near-optimal performance on the existing Frontier hardware. The framework is transferable to other liquid-cooled high-performance computing plants.
\end{abstract}


\section{Introduction}\label{sec:intro}

Data center electricity consumption has emerged as one of the most rapidly growing loads on the global power system. Worldwide data center demand reached 415~TWh in 2024, approximately 1.5\% of total electricity use, and is projected to exceed 945~TWh by 2030~\cite{IEA2025}. In the United States, data centers consumed 176~TWh in 2023, about 4.4\% of national electricity, with projections ranging from 325 to 580~TWh by 2028~\cite{Shehabi2024}. Cooling infrastructure accounts for 30 to 40\% of total facility electricity consumption~\cite{Zhang2021, Ebrahimi2014}, making it the single largest controllable load in most data centers. Despite a decade of advances in efficient server hardware and air-side economization, the global average Power Usage Effectiveness has remained between 1.55 and 1.59 since 2020~\cite{UptimeInstitute2024}, indicating that incremental improvements to existing cooling systems are approaching diminishing returns and that more systematic optimization approaches are needed.

High-performance computing (HPC) facilities experience this challenge at an amplified scale. Exascale supercomputers such as Frontier at Oak Ridge National Laboratory (ORNL) operate at power levels of 8 to 30~MW, employ 100\% direct liquid cooling with variable-speed pumps, and exhibit rapid thermal transients driven by workload dynamics~\cite{Sun2024, Karimi2024}. These systems differ structurally from enterprise data centers in three important ways. First, liquid cooling loops respond to load changes on time scales of minutes rather than the seconds typical of forced-air cooling, introducing tight coupling between thermal inertia and actuator rate limits. Second, equipment thermal constraints such as the 45~$^{\circ}$C coolant return temperature limit on direct-to-chip cold plates are operationally binding during high-load periods, leaving less margin for aggressive setpoint optimization. Third, the cooling plant architecture at facility scale is organized into multiple parallel subloops, each with independent pumps and heat exchangers, which creates a combinatorial design space that does not arise in smaller installations.

Despite the scale of energy involved, systematic cooling plant optimization for HPC facilities remains underdeveloped. Existing work on the Frontier system has focused on operational data collection and dataset release~\cite{Sun2024}, system-wide power management~\cite{Karimi2024}, thermal stability during load transients~\cite{Grant2026}, waste heat recovery integration~\cite{Wang2024P}, and data-driven diagnosis of cooling inefficiency~\cite{Jadhav2026ML}. These studies establish the physical and data foundation for optimization but do not prescribe control or design actions. In our recent work~\cite{Jadhav2026ML}, a three-stage machine learning framework was applied to the Frontier 2023 operational dataset to identify cooling energy waste, quantify approximately 85~MWh of annual inefficiency, and demonstrate that up to 96\% of the identified excess could be recovered through guardrail-constrained counterfactual setpoint adjustments. That work established the existence and magnitude of cooling energy waste at Frontier but did not construct a physics-based plant model and did not address the plant design question of how the 25 CDUs should be allocated across the parallel subloops.

A complementary line of work addresses cooling plant optimization through physics-based digital twins. The Modelica language and its Buildings Library~\cite{Wetter2014} have become the dominant open-source platform for equation-based modeling of HVAC and cooling systems, and Modelica models have been developed for air-cooled data centers~\cite{Fu2019a, Fu2019b}, district cooling systems~\cite{Hinkelman2022}, chiller plants with water-side economization~\cite{Fan2021}, and comparative co-design of cooling control strategies~\cite{Grahovac2023}. For HPC specifically, the Exascale Digital Twin (ExaDigiT) project at ORNL developed a digital twin of liquid-cooled supercomputers using the TRANSFORM library in Modelica ~\cite{Brewer2024, Kumar2024, Greenwood2024}. However, the ExaDigiT effort prioritized verification and validation over systematic optimization, and the question of how to use a validated digital twin as the basis for design-space exploration was left to subsequent work.

Our own recent contribution in this area~\cite{Jadhav2026Stage1} constructed a Modelica-based digital twin of the Frontier hot-temperature water (HTW) cooling system, validated it through one full calendar year of 10-minute operational data following ASHRAE Guideline~14 \cite{ASHRAE2014}, and used the validated twin to evaluate three progressively constrained control strategies: a flow-only optimization achieving 20.4\% total cooling energy savings, an unconstrained joint optimization of flow rate and supply temperature reaching 30.1\%, and a ramp-constrained joint optimization retaining 27.8\% (corresponding to a recovery ratio of 92.4\% relative to the unconstrained optimum). That work demonstrated that a validated digital twin can serve as an optimization testbed for HPC cooling control, and it quantified the gap between theoretical and deployable control strategies. However, it treated the cooling plant topology as fixed, accepting the existing Frontier design of three active parallel subloops with 14, 6, and 5 CDUs each, while not considering the plant design for co-design optimization.

This research focuses on the co-design optimization for data center. We established a validated digital twin of the Frontier cooling system and provided operational control strategies to optimize the plant configuration to lower annual cooling energy. The plant design space contains three coupled classes of decisions. The first is the number of parallel subloops, $K$, which determines how the total thermal load is distributed across independent hydraulic branches. The second is the allocation of the 25 CDUs across the $K$ subloops, which is an integer partition problem with 611 feasible configurations for $K$ between 2 and 6. The third is the flow fraction allocation $f_k$ across the subloops, which can be fixed at design time through hydraulic balancing valves or optimized dynamically through independent per-subloop flow control. These three classes of decisions interact through the per-subloop thermal constraint and the shared pump and cooling tower energy cost, producing a mixed-integer nonlinear optimization problem whose structure and solution have not been previously analyzed for the exascale HPC cooling plant.

A related body of work addresses plant-level optimization in chiller plants and district cooling systems, where decisions about equipment sizing, sequencing, and flow distribution have been studied using mixed-integer programming~\cite{Lu2019, Huang2020}, metaheuristic search~\cite{Afroz2018}, and surrogate-assisted optimization~\cite{Ma2012}. These methods are effective when the plant architecture is known a priori and the decisions reduce to equipment selection and operating schedules. They do not directly address the partition and flow allocation problem arising in multi-subloop HPC cooling plants, where the integer decision space is structured by the CDU count and the flow fractions interact nonlinearly with the per-subloop thermal intensity. The closest analog in the thermal systems literature is the heat exchanger network synthesis problem~\cite{Furman2002, Escobar2013}, which similarly combines integer topology decisions with continuous flow and temperature optimization. The framework developed in the present paper applies a similar decomposition approach (enumeration over integer decisions combined with gradient-based optimization of continuous decisions) to the specific structure of the HPC cooling plant, where the integer space is small enough to enumerate exhaustively, and the continuous subproblem admits analytical gradients.

The contributions of this paper are:

\begin{enumerate}
    \item A three-layer optimization framework is developed that decomposes the cooling plant design problem into the integer partition of CDUs across subloops (Layer~1), the continuous flow fraction allocation across subloops (Layer~2), and the per-timestep co-design optimization of total flow rate and supply temperature subject to thermal safety constraints (Layer~3). The framework is solved by exhaustive enumeration of the 611 feasible partitions combined with sequential least-squares programming (SLSQP) on the continuous subproblem, providing a global optimality certificate over the discrete design space.

    \item The framework is applied to the Frontier cooling plant using a reduced-order physics model calibrated through the Stage~1 Modelica digital twin~\cite{Jadhav2026Stage1}, and all 611 partitions are evaluated across the full 2023 operational dataset~\cite{Sun2024} under both balanced and worst-case CDU-to-subloop assignments.

    \item A design decision hierarchy is established in which the operational strategy dominates the number of subloops, which in turn dominates the CDU-to-subloop allocation. The optimal design over the enumerated surrogate problem is a two-subloop plant with partition $(19, 6)$, which achieves 35.48\% total cooling energy savings, only 0.18\% above the current three-subloop Frontier design of $(14, 6, 5)$ at 35.30\%.

    \item The analysis shows that flow fraction optimization compensates for any feasible CDU-to-subloop assignment, reducing the worst-case design sensitivity by 93\%. This identifies flow fraction control as a source of design robustness and provides a direct and low-cost recommendation for the Frontier facility and a transferable framework for other liquid-cooled HPC plants.
\end{enumerate}

The remainder of the paper is organized as follows. Section~\ref{sec:methodology} presents the three-layer optimization framework, the reduced-order physics model, and the computational implementation. Section~\ref{sec:results} presents the optimization results across all 611 partitions, including the optimal subloop count, strategy comparison, flow fraction optimization, and the sensitivity analyses. Section~\ref{sec:discussion} discusses the practical implications for plant design and commissioning, the substitutability between flow fraction optimization and workload scheduling, and the limitations and future work. Section~\ref{sec:conclusion} concludes.

\section{Methodology}\label{sec:methodology}

This section presents the three-layer optimization framework used to identify the energy-optimal cooling plant design for the Frontier supercomputer. Section~\ref{sec:method:overview} introduces the framework as a whole. Section~\ref{sec:method:rom} describes the reduced-order physics model used to evaluate each candidate design. Sections~\ref{sec:method:layer1}, \ref{sec:method:layer2}, and \ref{sec:method:layer3} present the three optimization layers. Section~\ref{sec:method:strategies} defines the three operational strategies that are compared in Section~\ref{sec:results}. Section~\ref{sec:method:implementation} summarizes the computational implementation.

\subsection{Three-Layer Framework Overview}\label{sec:method:overview}

\begin{figure*}[t]
\centering
\includegraphics[width=\textwidth]{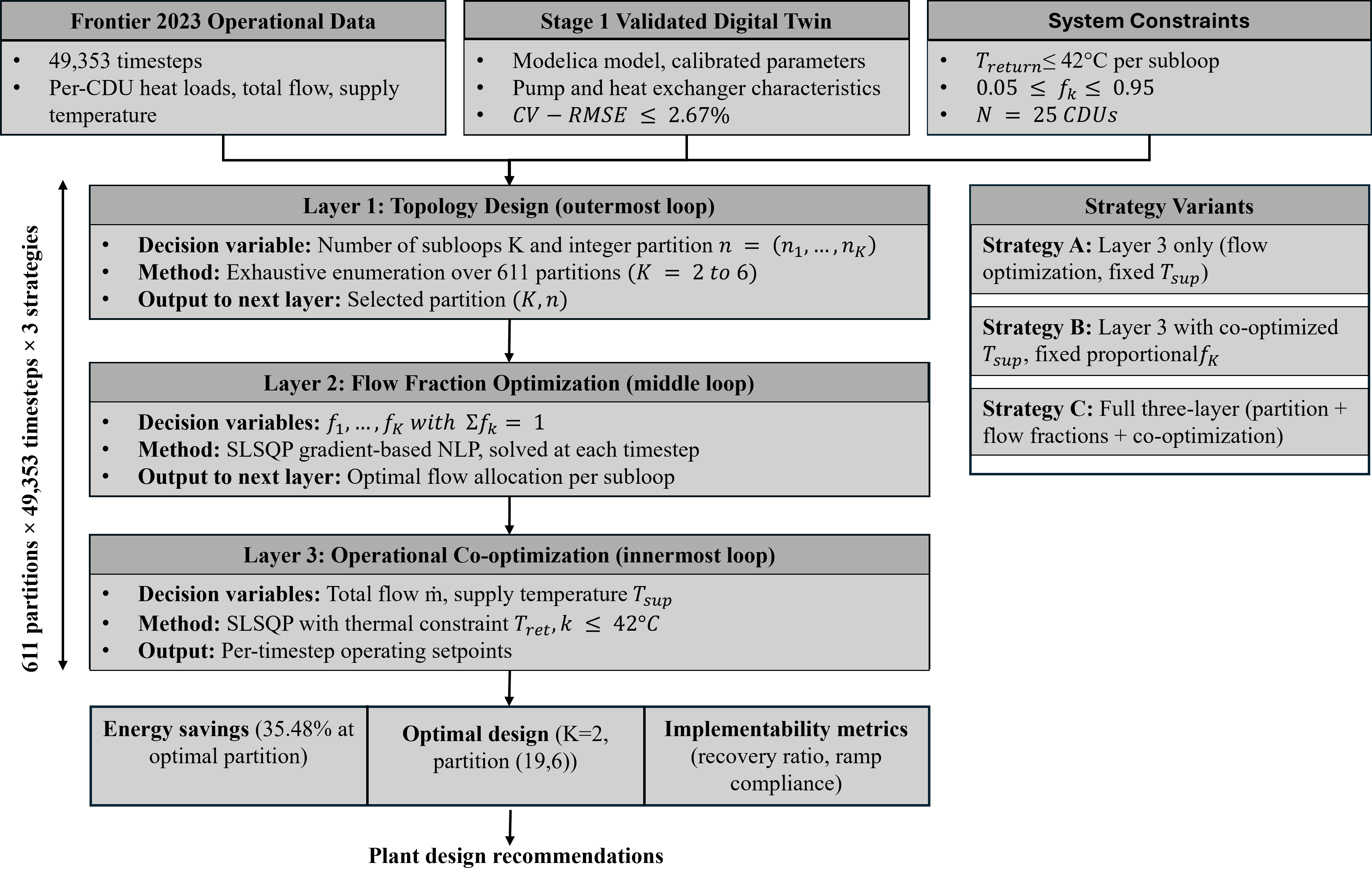}
\caption{Three-layer optimization framework for the Frontier cooling system. Layer~1 selects the integer partition of CDUs into subloops, Layer~2 optimizes the continuous flow fractions, and Layer~3 co-optimizes the total flow rate and supply temperature setpoint at each timestep. Strategy variants A, B, and C are defined as progressively richer instances of the framework.}
\label{fig:framework}
\end{figure*}

The cooling system co-design optimization problem is decomposed into three nested layers, each addressing a distinct class of decision variables. Figure~\ref{fig:framework} presents the overall structure. Layer~1 (topology design) determines the integer partition of $N=25$ CDUs into $K$ subloops. Layer~2 (flow fraction optimization) determines the continuous flow allocation among the $K$ subloops. Layer~3 (operational co-design optimization) determines the total flow rate and the supply temperature setpoint at each timestep, subject to the per-subloop thermal safety constraint. The three layers together form a mixed-integer nonlinear program in which the integer decisions of Layer~1 are resolved by exhaustive enumeration over all 611 feasible partitions, and the continuous decisions of Layers~2 and~3 are solved by SLSQP at each of the 49{,}353 operating timesteps in the 2023 dataset~\cite{Sun2024}.

The framework takes three inputs: the Frontier 2023 operational dataset~\cite{Sun2024} containing 49{,}353 timesteps of measured per-CDU heat loads, total flow rates, and supply temperatures; the validated Stage~1 Modelica digital twin~\cite{Jadhav2026Stage1}, which provides calibrated pump and heat exchanger characteristics with coefficient of variation of the root mean squared error (CV-RMSE) between 1.96 and 2.67\% compared to ASHRAE Guideline~14 \cite{ASHRAE2014}; and the system constraints, namely the per-subloop return temperature limit $T_{\mathrm{ret},k} \leq 42^{\circ}$C, the bounds on subloop flow fractions $0.05 \leq f_k \leq 0.95$, and the total CDU count $N=25$. The framework produces three outputs: the optimal partition $(K^{*}, \mathbf{n}^{*})$, the corresponding annual energy savings, and the implementability metrics that quantify how the optimized plant compares with the baseline operation.

The decomposition into three layers is justified by the structure of the problem. The integer partition variable in Layer~1 has finite cardinality (611 partitions for $K \in \{2, 3, 4, 5, 6\}$), so exhaustive enumeration is both tractable and provably optimal over the discrete design space, providing a global optimality certificate that branch-and-bound or outer-approximation methods cannot guarantee for nonconvex thermal-hydraulic models. The continuous variables in Layers~2 and~3 enter the objective and constraint functions through smooth nonlinear expressions (the pump affinity law, the cooling tower fan power correlation, and the per-subloop energy balance), making them well-suited to gradient-based optimization. SLSQP is selected because it handles inequality constraints directly, converges quadratically near the optimum, and has been used successfully in the Stage~1 control optimization with the same physics model.

\subsection{Reduced-Order Physics Model}\label{sec:method:rom}

The co-design optimization framework requires evaluating thousands of candidate designs across tens of thousands of timesteps each, which is computationally infeasible with the full Modelica digital twin. A reduced-order Python model is therefore developed and calibrated through the Stage~1 Modelica outputs. The reduced-order model uses three sub-models: a pump affinity law for total HTW pump power, a fan power correlation for the cooling tower, and a per-subloop energy balance for the return temperature constraint.

The total HTW pump power $P_{\mathrm{pump}}$ is computed from the affinity law:
\begin{equation}
P_{\mathrm{pump}}(\dot{m}) = P_{\mathrm{pump,nom}} \left( \frac{\dot{m}}{\dot{m}_{\mathrm{nom}}} \right)^{n_p}
\label{eq:pump_power}
\end{equation}
where $\dot{m}$ is the total HTW mass flow rate, $\dot{m}_{\mathrm{nom}} = 190$~kg/s is the nominal flow rate at the design point, $P_{\mathrm{pump,nom}} = 17.18$~kW is the calibrated nominal pump power, and $n_p$ is the pump power exponent. A nominal value of $n_p = 3$ is used following the cubic affinity law for variable-speed centrifugal pumps; sensitivity to $n_p$ is examined in Section~\ref{sec:results}. Equation~\eqref{eq:pump_power} represents the aggregate hydraulic power delivered by the three active variable-speed pumps in the Frontier HTW system; each pump is independently controlled and is modeled implicitly through the per-subloop flow fractions in Layer~2.

The cooling tower fan power $P_{\mathrm{ct}}$ is modeled as a function of the heat rejection rate and the approach temperature:
\begin{equation}
P_{\mathrm{ct}}(\dot{m}, T_{\mathrm{sup}}) = P_{\mathrm{ct,nom}} \, \frac{Q_{\mathrm{rej}}}{Q_{\mathrm{rej,nom}}} \left( \frac{\Delta T_{\mathrm{app,nom}}}{\Delta T_{\mathrm{app}}} \right)
\label{eq:ct_power}
\end{equation}
where $Q_{\mathrm{rej}}$ is the total heat rejected by the cooling tower at the current operating point, $Q_{\mathrm{rej,nom}} = 9170.7$~kW is the nominal heat rejection rate, $\Delta T_{\mathrm{app}}$ is the cooling tower approach temperature defined as the difference between the CTW supply temperature and the ambient wet-bulb temperature, $\Delta T_{\mathrm{app,nom}} = 4.0^{\circ}$C is the nominal approach, and $P_{\mathrm{ct,nom}} = 950.13$~kW is the calibrated nominal fan power. The approach temperature at the current operating point is computed as $\Delta T_{\mathrm{app}} = \Delta T_{\mathrm{app,base}} + (T_{\mathrm{sup}} - T_{\mathrm{sup}}^{\mathrm{base}})$, where $\Delta T_{\mathrm{app,base}} = 23.5^{\circ}$C is the cooling tower approach temperature at the measured baseline operating point and $T_{\mathrm{sup}}^{\mathrm{base}}$ is the corresponding baseline supply temperature. Because the ambient wet-bulb temperature varies slowly relative to the supply temperature setpoint, raising the supply temperature above its baseline value widens the tower approach by an equal amount, which is the physical content captured by this expression. Equation~\eqref{eq:ct_power} is an empirical correlation calibrated through the Stage~1 digital twin rather than a first-principles fan-affinity model; it is fitted at the baseline operating point and applied over the supply-temperature envelope of 27 to 35$^{\circ}$C explored by the optimized strategies, and it is not extrapolated beyond that range. It reproduces the established trade-off between supply temperature setpoint and cooling tower fan energy: a higher supply temperature yields a wider approach and reduces the fan duty required to reject the same total heat load.

The per-subloop return temperature is computed from the steady-state energy balance:
\begin{equation}
T_{\mathrm{ret},k} = T_{\mathrm{sup}} + \frac{Q_k}{f_k \, \dot{m} \, c_p}, \quad k = 1, \ldots, K
\label{eq:tret}
\end{equation}
where $Q_k$ is the heat load assigned to subloop $k$, $f_k$ is the flow fraction in subloop $k$ (with $\sum_k f_k = 1$), $T_{\mathrm{sup}}$ is the supply temperature setpoint, and $c_p = 3500$~J/(kg$\cdot$K) is the specific heat capacity of the 50/50 ethylene glycol-water mixture. The thermal safety constraint requires $T_{\mathrm{ret},k} \leq T_{\mathrm{limit}} = 42^{\circ}$C for all $k$, with $T_{\mathrm{limit}}$ set $3^{\circ}$C below the equipment specification of $45^{\circ}$C to provide a safety margin. The constraint is imposed on the subloop return temperature, which, under the lumped energy balance, governs the aggregate cold-plate loop and is the variable reported by the facility instrumentation. Nonuniform heat loading across individual racks or chips within a subloop is not resolved by this reduced-order balance; the $3^{\circ}$C margin below the equipment limit is retained in part to accommodate such within-subloop nonuniformity, and a finer-grained thermal model is identified as future work in Section~\ref{sec:discuss:limitations}.

The total cooling plant power at any operating point is the sum of the pump and fan contributions:
\begin{equation}
P_{\mathrm{total}}(\dot{m}, T_{\mathrm{sup}}, \mathbf{f}) = P_{\mathrm{pump}}(\dot{m}) + P_{\mathrm{ct}}(\dot{m}, T_{\mathrm{sup}})
\label{eq:ptotal}
\end{equation}
which is the objective function minimized at each timestep. The reduced-order model is calibrated so that, when evaluated under the baseline operating conditions of the 2023 dataset, the annual cooling energy reproduces the Stage~1 baseline of 488{,}857~kWh for the pumps and 1{,}324{,}797~kWh for the cooling tower fans, matching the Stage~1 paper to within 0.1\%.

\subsection{Layer 1: Plant Topology Optimization}\label{sec:method:layer1}

Layer~1 determines the integer partition of $N=25$ CDUs into $K$ subloops. A partition is an ordered tuple $\mathbf{n} = (n_1, n_2, \ldots, n_K)$ with $n_k \geq 1$ for all $k$, $\sum_k n_k = N$, and $n_1 \geq n_2 \geq \ldots \geq n_K$. The ordering convention $n_1 \geq n_2 \geq \ldots \geq n_K$ removes permutation symmetry and ensures that each distinct topology is counted exactly once.

The set of feasible partitions is $\mathcal{P}_K$ for each $K$, and the full design space is $\mathcal{P} = \bigcup_{K=2}^{6} \mathcal{P}_K$. The cardinalities are $|\mathcal{P}_2| = 12$, $|\mathcal{P}_3| = 52$, $|\mathcal{P}_4| = 120$, $|\mathcal{P}_5| = 192$, and $|\mathcal{P}_6| = 235$, giving a total of $|\mathcal{P}| = 611$ distinct designs. The range of subloop counts $K \in \{2, \ldots, 6\}$ is bounded by physical and architectural constraints, as discussed in Section~\ref{sec:results}.

For each partition $\mathbf{n} \in \mathcal{P}$, the optimizer must also assign the 25 CDUs to the $K$ subloops in a way that determines the per-subloop heat load $Q_k$. Two assignment policies are evaluated. The balanced assignment groups CDUs by sorted heat load such that each subloop receives a representative mix of high-load and low-load CDUs, producing per-subloop heat loads that are approximately proportional to $n_k$. The worst-case assignment groups all of the highest-load CDUs into a single subloop, producing the most thermally imbalanced configuration consistent with the partition. Reporting both cases bounds the achievable performance under different commissioning practices.

Layer~1 is solved by exhaustive enumeration. Each of the 611 partitions is evaluated independently, and the partition that yields the minimum annual cooling energy under the chosen strategy is selected as the global optimum. Exhaustive enumeration is both tractable (the inner SLSQP solve completes in approximately 10~ms per timestep, so the full annual evaluation of one partition completes in about 8 minutes on a single CPU core) and provably optimal over the discrete design space.

\subsection{Layer 2: Flow Fraction Optimization}\label{sec:method:layer2}

Layer~2 determines the continuous flow fractions $\mathbf{f} = (f_1, \ldots, f_K)$ that allocate the total HTW flow $\dot{m}$ across the $K$ subloops. The flow fractions satisfy the mass balance $\sum_{k=1}^{K} f_k = 1$ and the per-subloop bounds $0.05 \leq f_k \leq 0.95$, which prevent the optimizer from assigning negligible flow to any active subloop and ensure that the reduced-order energy balance remains numerically well-conditioned.

For each timestep $t$ and each candidate partition $\mathbf{n}$, the Layer~2 problem solves for $\mathbf{f}(t)$ together with the Layer~3 control variables. The flow fractions are not held constant across timesteps; instead, they are re-optimized at each timestep in response to the instantaneous per-CDU heat load distribution. This allows the framework to compensate for the natural variation in workload that occurs over the year and to direct more flow to subloops with higher instantaneous thermal load.

For comparison, a fixed proportional baseline is also evaluated in which $f_k = n_k / N$ for every timestep. The fixed proportional case represents the minimum-modification deployment scenario in which the flow fractions are determined at design time by hydraulic balancing valves and remain constant during operation. The benefit of dynamic Layer~2 optimization over this fixed baseline is reported in Section~\ref{sec:results}.

\subsection{Layer 3: Control Co-Design Optimization}\label{sec:method:layer3}

Layer~3 determines the total HTW mass flow rate $\dot{m}(t)$ and the supply temperature setpoint $T_{\mathrm{sup}}(t)$ at each timestep. These are the two operational degrees of freedom available to the supervisory controller of the existing Frontier cooling plant. The Layer~3 optimization problem at timestep $t$ is:
\begin{equation}
\min_{\dot{m},\, T_{\mathrm{sup}},\, \mathbf{f}} \quad P_{\mathrm{total}}(\dot{m}, T_{\mathrm{sup}}, \mathbf{f})
\label{eq:objective}
\end{equation}
subject to
\begin{align}
T_{\mathrm{sup}} + \frac{Q_k(t)}{f_k \, \dot{m} \, c_p} &\leq T_{\mathrm{limit}}, \quad k = 1, \ldots, K \label{eq:c1}\\
\dot{m}_{\min} \leq \dot{m} &\leq \dot{m}_{\max} \label{eq:c2}\\
T_{\mathrm{sup,min}} \leq T_{\mathrm{sup}} &\leq T_{\mathrm{sup,max}} \label{eq:c3}\\
0.05 \leq f_k &\leq 0.95, \quad k = 1, \ldots, K \label{eq:c4}\\
\sum_{k=1}^{K} f_k &= 1 \label{eq:c5}
\end{align}
where the physical bounds are $\dot{m}_{\min} = 30$~kg/s, $\dot{m}_{\max} = 420$~kg/s, $T_{\mathrm{sup,min}} = 10^{\circ}$C, and $T_{\mathrm{sup,max}} = 35^{\circ}$C. The constraint set is the intersection of the per-subloop thermal safety constraint~\eqref{eq:c1}, the actuator bounds~\eqref{eq:c2} and~\eqref{eq:c3}, the flow fraction bounds~\eqref{eq:c4}, and the mass balance~\eqref{eq:c5}.

The problem is solved by SLSQP with analytical gradients of the objective and the inequality constraints. At each timestep, the previous-timestep solution is used as the initial guess to accelerate convergence. Each timestep is solved as an independent steady-state constrained optimization; the warm start transfers only the initial guess and not any dynamic state. The procedure is therefore not a model predictive controller. It carries no prediction horizon, does not propagate thermal or hydraulic dynamics between timesteps, and does not anticipate future load. It is best understood as a quasi-steady evaluation of the achievable operating point at each measured load condition, which is appropriate here because the cooling loop time constants are short relative to the 10-minute telemetry interval and the analysis targets annual energy rather than transient control performance. The annual total energy under a given partition is obtained by summing the per-timestep optimal $P_{\mathrm{total}}$ over all 49{,}353 timesteps and multiplying by the 600~s timestep length:
\begin{equation}
E_{\mathrm{annual}}(\mathbf{n}) = \sum_{t=1}^{49{,}353} P_{\mathrm{total}}^{*}(t; \mathbf{n}) \cdot \Delta t
\label{eq:annual}
\end{equation}
where $P_{\mathrm{total}}^{*}(t; \mathbf{n})$ is the optimal value of the objective at timestep $t$ for partition $\mathbf{n}$.

\subsection{Operational Strategies}\label{sec:method:strategies}

Three operational strategies are defined as progressively richer instances of the three-layer framework. Each strategy is evaluated independently for every partition in $\mathcal{P}$, and the resulting annual energy values are compared in Section~\ref{sec:results}.

Strategy~A (flow-only optimization) activates only Layer~3 with $T_{\mathrm{sup}}$ held at the measured baseline value $T_{\mathrm{sup}}^{\mathrm{base}}(t)$ at every timestep, and uses fixed proportional flow fractions $f_k = n_k / N$. The optimizer reduces the total flow rate $\dot{m}(t)$ to the minimum value consistent with the thermal safety constraint, exploiting the cubic dependence of pump power on flow rate without modifying any other operating variable. Strategy~A represents the most conservative deployment, requiring no setpoint changes and no flow control hardware modifications.

Strategy~B (co-design optimization with fixed flow fractions) activates Layer~3 with both $\dot{m}$ and $T_{\mathrm{sup}}$ as decision variables, while still using fixed proportional flow fractions $f_k = n_k / N$. The supply temperature setpoint is allowed to rise above the measured baseline to reduce cooling tower fan energy, subject to the thermal safety constraint. Strategy~B represents the upper bound of what can be achieved by supervisory setpoint optimization on a plant with fixed proportional flow distribution.

Strategy~C (full three-layer optimization) activates all three layers. The flow fractions $\mathbf{f}$, the total flow rate $\dot{m}$, and the supply temperature setpoint $T_{\mathrm{sup}}$ are all decision variables at each timestep. Strategy~C represents the energy-optimal deployment under the assumption that each subloop has independently controllable flow, which is the case for the Frontier HTW pumps already in service.

\subsection{Computational Implementation}\label{sec:method:implementation}

The three-layer framework is implemented in Python 3.11 using NumPy for the reduced-order model evaluation and SciPy~\cite{Virtanen2020} for the SLSQP optimization. The 611 partitions are evaluated in parallel using the Python multiprocessing module across 32 CPU cores on a Linux workstation. The total wall-clock time for the full sweep, including all three strategies, is approximately 15 hours.

The reduced-order model is calibrated using the Stage~1 Modelica digital twin~\cite{Jadhav2026Stage1} by matching the annual baseline energy consumption to within 0.1\%. The calibrated parameter values are $P_{\mathrm{pump,nom}} = 17.18$~kW, $P_{\mathrm{ct,nom}} = 950.13$~kW, $\dot{m}_{\mathrm{nom}} = 190$~kg/s, $Q_{\mathrm{rej,nom}} = 9170.7$~kW, $\Delta T_{\mathrm{app,nom}} = 4.0^{\circ}$C, $\Delta T_{\mathrm{app,base}} = 23.5^{\circ}$C, $c_p = 3500$~J/(kg$\cdot$K), and $T_{\mathrm{limit}} = 42^{\circ}$C. The dataset is the publicly available Frontier 2023 telemetry record~\cite{Sun2024}, filtered to remove maintenance shutdowns and sensor anomalies, leaving 49{,}353 valid timesteps at 10-minute intervals.

The SLSQP solver is configured with a function tolerance of $10^{-8}$ and a maximum of 200 iterations per timestep. Convergence is achieved within 10~ms per timestep on average, with worst-case convergence times of 50~ms during periods of rapid load transients. Failed timesteps (defined as solver iterations exceeding the maximum without satisfying the convergence tolerance) are handled by falling back to the previous-timestep solution; fewer than 0.1\% of timesteps require this fallback across all partitions and strategies.

\section{Results}\label{sec:results}

This section presents the results of the three-layer optimization framework applied to the Frontier supercomputer cooling plant. All 611 integer partitions of $N=25$ CDUs into $K=2$ through $K=6$ subloops are evaluated using the full 2023 operational dataset \cite{Sun2024} containing 49{,}353 timesteps at 10-minute intervals. Each partition is assessed under both balanced and worst-case CDU-to-subloop assignments, and under all three operational strategies: flow-only optimization (Strategy~A), joint flow and supply temperature co-design optimization with proportional flow fractions (Strategy~B), and joint co-design optimization with SLSQP-optimized flow fractions (Strategy~C).

The range of subloop counts considered ($K=2$ to $K=6$) is bounded by physical and architectural constraints. The lower bound $K=1$ is excluded because a single-subloop configuration collapses the parallel topology into a single hydraulic branch, eliminating the distribution of heat load across independent flow paths that the three-loop architecture is designed to exploit. In this degenerate case, the CDU-to-subloop allocation decision no longer exists, and the optimization reduces to the Stage~1 single-loop control problem already addressed in our prior work. The upper bound $K=6$ reflects the practical constraint that each additional subloop requires a dedicated plate heat exchanger, variable-speed pump, and instrumentation set; beyond six subloops, the fixed infrastructure cost grows faster than the achievable energy benefit, and partitions with $n_k=1$ CDU per subloop become common, leading to heat exchanger effectiveness degradation below 0.5. These bounds are consistent with the range of liquid cooling architectures observed in leadership-class HPC facilities \cite{Sun2024}.

\subsection{Optimal Subloop Count}\label{sec:results:Kopt}

\begin{figure*}[t]
\centering
\includegraphics[width=\textwidth]{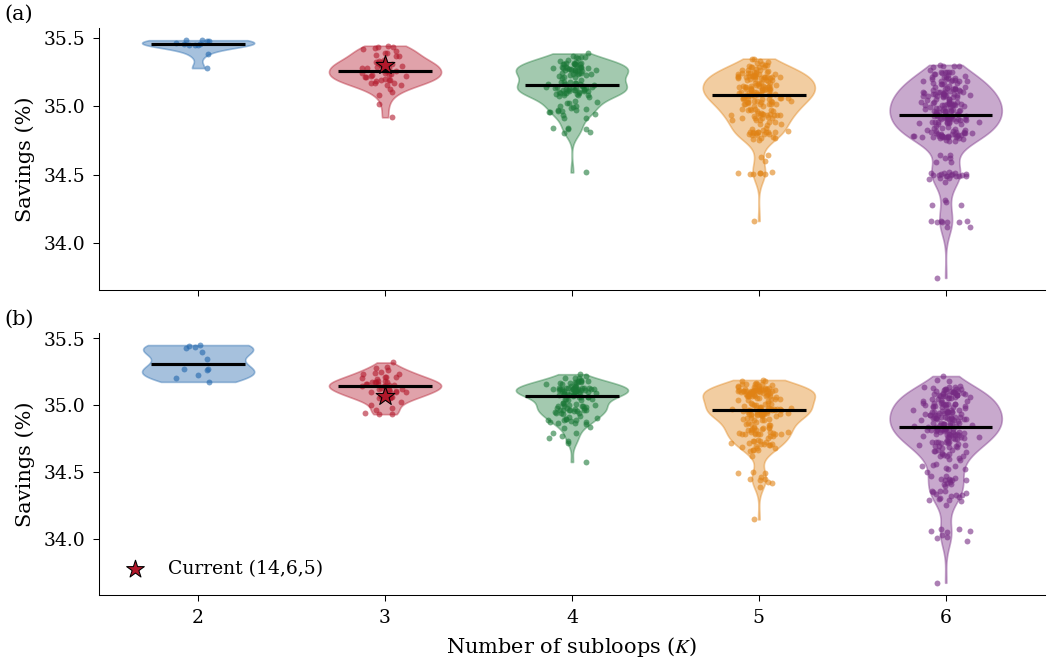}
\caption{Strategy~C savings as a function of the number of subloops ($K$) under (a) balanced and (b) worst-case CDU assignment. The star marker denotes the current Frontier design $(14, 6, 5)$.}
\label{fig:violin}
\end{figure*}

Figure~\ref{fig:violin} presents the violin plots of energy savings of Strategy~C across all 611 partitions, grouped by the number of subloops $K$. Under both balanced assignment (Fig.~\ref{fig:violin}(a)) and worst-case assignment (Fig.~\ref{fig:violin}(b)), $K=2$ achieves the highest mean savings and the narrowest distribution. The mean savings decrease monotonically as $K$ grows, from 35.44\% at $K=2$ (range: 35.28 - 35.48\%) to 34.89\% at $K=6$ (range: 33.75 - 35.30\%). The width of the savings distribution grows substantially with $K$: the 0.20\% range observed at $K=2$ broadens to a 1.55\% range at $K=6$, indicating that the CDU-to-subloop allocation decision becomes progressively more consequential as the number of subloops increases. The current Frontier design, a $K=3$ partition of $(14, 6, 5)$, is marked with a star symbol in Fig.~\ref{fig:violin}(a) and achieves 35.30\% savings, placing it 0.18\% below the optimum at $K=2$ with partition $(19, 6)$, a margin shown in Section~\ref{sec:results:uncertainty} to lie within the uncertainty of the reduced-order surrogate.

The physical basis for the advantage of smaller $K$ deserves explanation, as it may initially appear counterintuitive. For a fixed total heat load $Q_{\mathrm{tot}}$ and total flow rate $\dot{m}$, the binding thermal constraint in the optimization is the maximum return temperature across subloops, which is governed by the single subloop carrying the highest heat intensity $w_k / f_k$. Applying Jensen's inequality to the convex function $g(x) = x$ over the distribution of subloop heat intensities shows that, for a fixed mean, the maximum of the distribution grows with the number of subloops when the heat load distribution is non-uniform. Fewer subloops yield fewer independent ways for imbalance to occur, and the maximum heat intensity across subloops is consequently smaller, relaxing the thermal constraint and permitting lower flow rates and higher supply temperature setpoints.

This explanation also clarifies why $K=1$ is not included in the analysis and does not represent a meaningful extrapolation of the trend. A single-subloop configuration ($K=1$) is not simply the limit of the observed trend because the parallel-loop architecture required for independent thermal management no longer exists. In the $K=1$ case, the entire heat load passes through a single branch, the CDU-to-subloop assignment problem disappears, and the optimization reduces to the global flow and temperature control problem solved in Stage~1. The monotonic trend toward smaller $K$ observed in the present results reflects a benefit within the class of parallel-subloop architectures, not a preference for eliminating the parallel topology altogether.

\subsection{Strategy Comparison}\label{sec:results:strategies}

\begin{figure}[h]
\centering
\includegraphics[width=0.5\columnwidth]{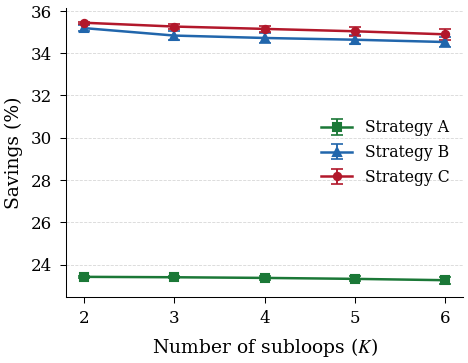}
\caption{Comparison of Strategies~A, B, and~C (best partition per $K$, balanced assignment). Error bars span the range across all partitions within each $K$.}
\label{fig:strategy_comparison}
\end{figure}

Figure~\ref{fig:strategy_comparison} compares the three operational strategies across all five values of $K$. Strategy~A (flow-only optimization) achieves approximately 23.4\% savings independent of $K$ and partition choice, because the pump affinity law depends only on the total flow rate, not on how CDUs are allocated among subloops. Strategies~B and~C both achieve substantially higher savings by exploiting supply temperature reset to reduce cooling tower fan energy. The marginal benefit of Strategy~C over Strategy~B is between 0.2 and 0.5\% depending on $K$, with the advantage increasing at higher $K$ where flow redistribution can better compensate for heat load imbalance across subloops.

Table~\ref{tab:best_per_K} summarizes the best partition for each $K$ under Strategy~C with balanced CDU assignment. The global optimum occurs at $K=2$ with partition $(19, 6)$, yielding 35.48\% savings. All five best-per-$K$ partitions achieve savings above 35.30\%, confirming that the framework is robust to the choice of $K$ provided the CDU allocation and flow fractions are optimized. The recovery ratio, defined as the Strategy~C savings divided by the Strategy~B savings for the same partition, exceeds 100\% for the best partition at every $K$, a result discussed further in Section~\ref{sec:results:recovery}.

\begin{table}[t]
\caption{Best partition per subloop count $K$ under balanced CDU assignment. $s_A$, $s_B$, and $s_C$ are the annual savings of the flow-only, fixed proportional co-design, and full co-design strategies; the recovery ratio is $s_C / s_B$ with Strategy~B using the fixed proportional split $f_k = n_k / N$.}
\label{tab:best_per_K}
\centering
\resizebox{\columnwidth}{!}{
\begin{tabular}{c l c c c c c}
\toprule
$K$ & Partition $\mathbf{n}$ & $|\mathcal{P}_K|$ & $s_A$ (\%) & $s_B$ (\%) & $s_C$ (\%) & Recovery (\%) \\
\midrule
2 & (19, 6)             &  12 & 23.43 & 35.06 & 35.48 & 101.19 \\
3 & (20, 3, 2)          &  52 & 23.43 & 32.53 & 35.44 & 108.94 \\
4 & (14, 5, 4, 2)       & 120 & 23.43 & 35.32 & 35.38 & 100.17 \\
5 & (15, 3, 3, 2, 2)    & 192 & 23.43 & 32.53 & 35.35 & 108.65 \\
6 & (14, 3, 2, 2, 2, 2) & 235 & 23.42 & 31.66 & 35.30 & 111.51 \\
\bottomrule
\end{tabular}
}
\end{table}

\subsection{Value of Flow Fraction Optimization}\label{sec:results:flowopt}

\begin{figure}[h]
\centering
\includegraphics[width=0.5\columnwidth]{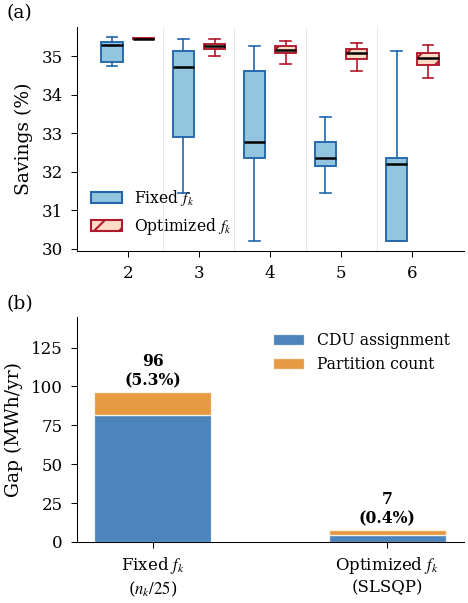}
\caption{(a)~Box plots of Strategy~C savings under fixed ($f_k = n_k / N$) and optimized flow fractions. (b)~Decomposition of the design savings spread into CDU assignment and partition count contributions.}
\label{fig:flow_opt}
\end{figure}

The central finding of this study is that optimizing the subloop flow fractions $f_k$ dramatically reduces the sensitivity of cooling plant performance to the CDU-to-subloop partition. Figure~\ref{fig:flow_opt} quantifies this effect. Figure~\ref{fig:flow_opt}(a) compares the distributions of Strategy~C savings under fixed proportional flow fractions ($f_k = n_k / N$) and SLSQP-optimized flow fractions, grouped by $K$. Under fixed $f_k$, the interquartile range spans 2 to 4\% at each $K$, reflecting the strong dependence of achievable savings on CDU allocation. Under optimized $f_k$, the distributions collapse to narrow bands of 0.2 to 0.5\%, and the median savings are nearly identical across all $K$.

Figure~\ref{fig:flow_opt}(b) presents the same information as a design savings improvement: the additional annual energy savings (MWh/yr) obtained by moving from the worst-case partition to the best-case partition across all 611 configurations. Under fixed $f_k$, this improvement is 96~MWh/yr, equivalent to 5.3\% of the baseline annual consumption. Flow fraction optimization reduces the spread between best and worst partitions to 7~MWh/yr (0.4\%), a 93\% reduction in the design sensitivity. The stacked bars decompose the improvement into two components: the CDU assignment effect (variability within a given $K$) and the partition count effect (variability between different values of $K$). Under fixed $f_k$, the CDU assignment effect dominates the design sensitivity. Under optimized $f_k$, both components shrink to near-negligible levels.

\begin{figure}[h]
\centering
\includegraphics[width=0.5\columnwidth]{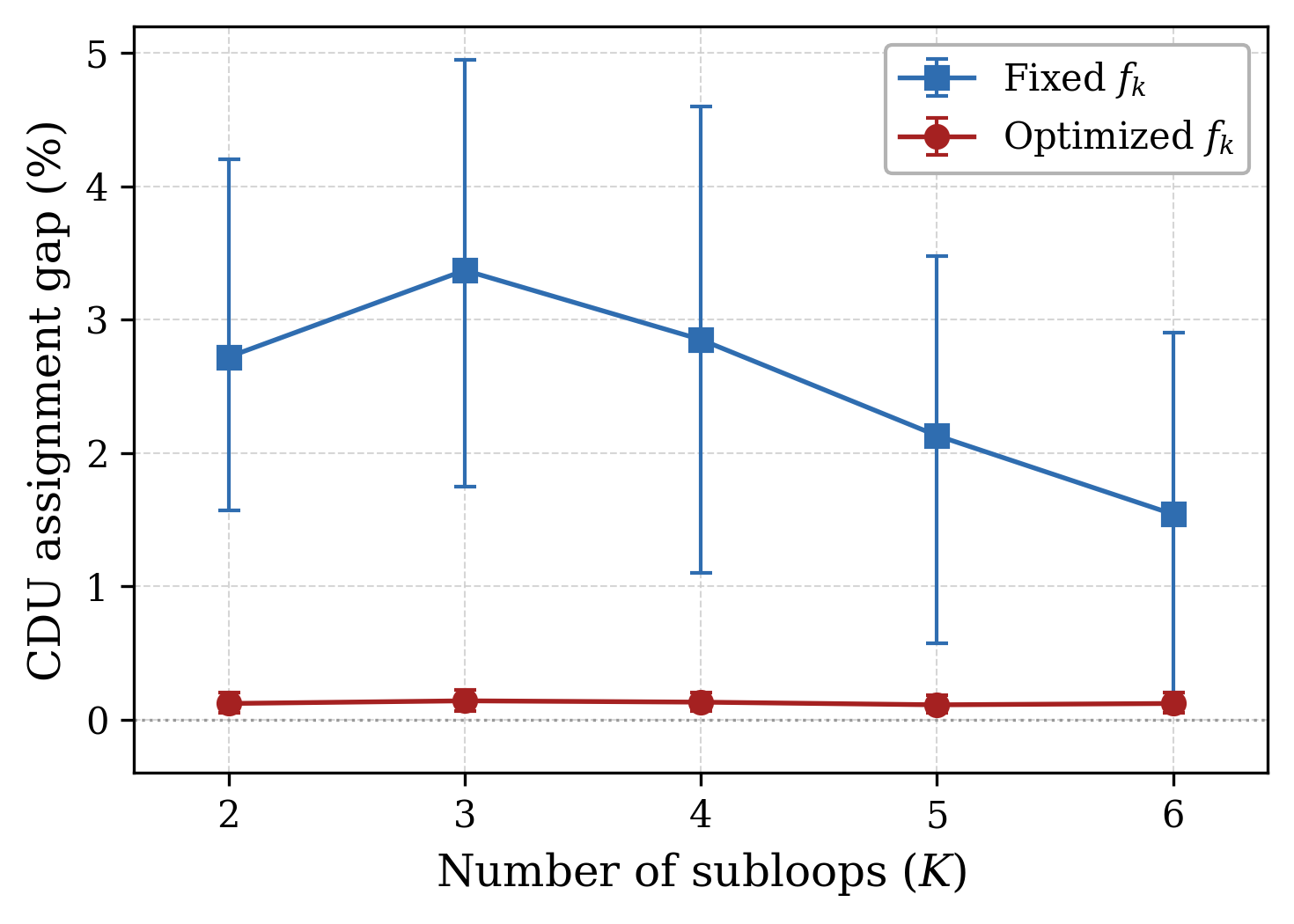}
\caption{CDU assignment gap (balanced minus worst-case savings) under fixed and optimized flow fractions, as a function of $K$.}
\label{fig:assignment_gap}
\end{figure}

Figure~\ref{fig:assignment_gap} presents the CDU assignment gap (the difference between best-case and worst-case CDU assignment savings for the same partition count $K$) as a function of $K$. Under fixed $f_k$, this gap averages 2 to 3\% with large variance across partitions, because an unfavorable CDU assignment produces subloops with systematically different heat intensities, and no corrective mechanism exists to compensate at the operational layer. Under optimized $f_k$, the assignment gap collapses to approximately 0.1\% with negligible variance across all $K$. The reason the optimized-$f_k$ curve is essentially flat is that the SLSQP optimizer is free to redirect flow from subloops with surplus cooling capacity toward subloops carrying heavier thermal loads at every timestep, equalizing the per-subloop thermal intensity $w_k / f_k$ regardless of how the CDUs were initially assigned to subloops. In other words, flow redistribution at the operational layer provides a corrective degree of freedom that absorbs the effect of any feasible CDU assignment, so the design-layer choice becomes nearly immaterial to the final energy outcome.

\subsection{Sensitivity to CDU Assignment}\label{sec:results:assignment}

\begin{figure}[h]
\centering
\includegraphics[width=0.5\columnwidth]{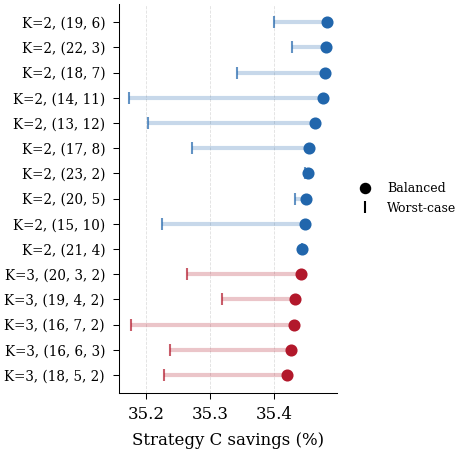}
\caption{Top-ranked partitions by Strategy~C savings (balanced assignment, circles) with worst-case assignment spread (bar endpoints). Blue indicates $K=2$ partitions; red indicates $K=3$ partitions.}
\label{fig:dumbbell}
\end{figure}

Figure~\ref{fig:dumbbell} examines the top-performing partitions individually, showing Strategy~C savings under both balanced (circle) and worst-case (bar endpoint) CDU assignments. All $K=2$ partitions, shown in blue, cluster at the top of the ranking with balanced-case savings above 35.2\% and a worst-case spread of 0.05 to 0.50\%. The $K=3$ partitions, shown in red, appear immediately below the $K=2$ group with comparable balanced-case savings but slightly wider worst-case spreads. This ranking is consistent with the monotonic trend observed in Fig.~\ref{fig:violin}: $K=2$ provides both the highest mean savings and the narrowest distribution. Even the worst-case performance of the best $K=2$ partition $(19, 6)$, at 35.40\%, exceeds the balanced-case performance of most $K=3$ partitions, reinforcing that $K=2$ is the globally preferred subloop count. The $K=3$ partitions in this figure are shown to document that the current Frontier topology (a $K=3$ design) remains near-optimal even though it is not globally optimal, which is the operationally relevant finding for a facility that cannot easily convert from three subloops to two subloops.

\subsection{Workload Equalization Sensitivity}\label{sec:results:workload}

\begin{figure}[h]
\centering
\includegraphics[width=0.5\columnwidth]{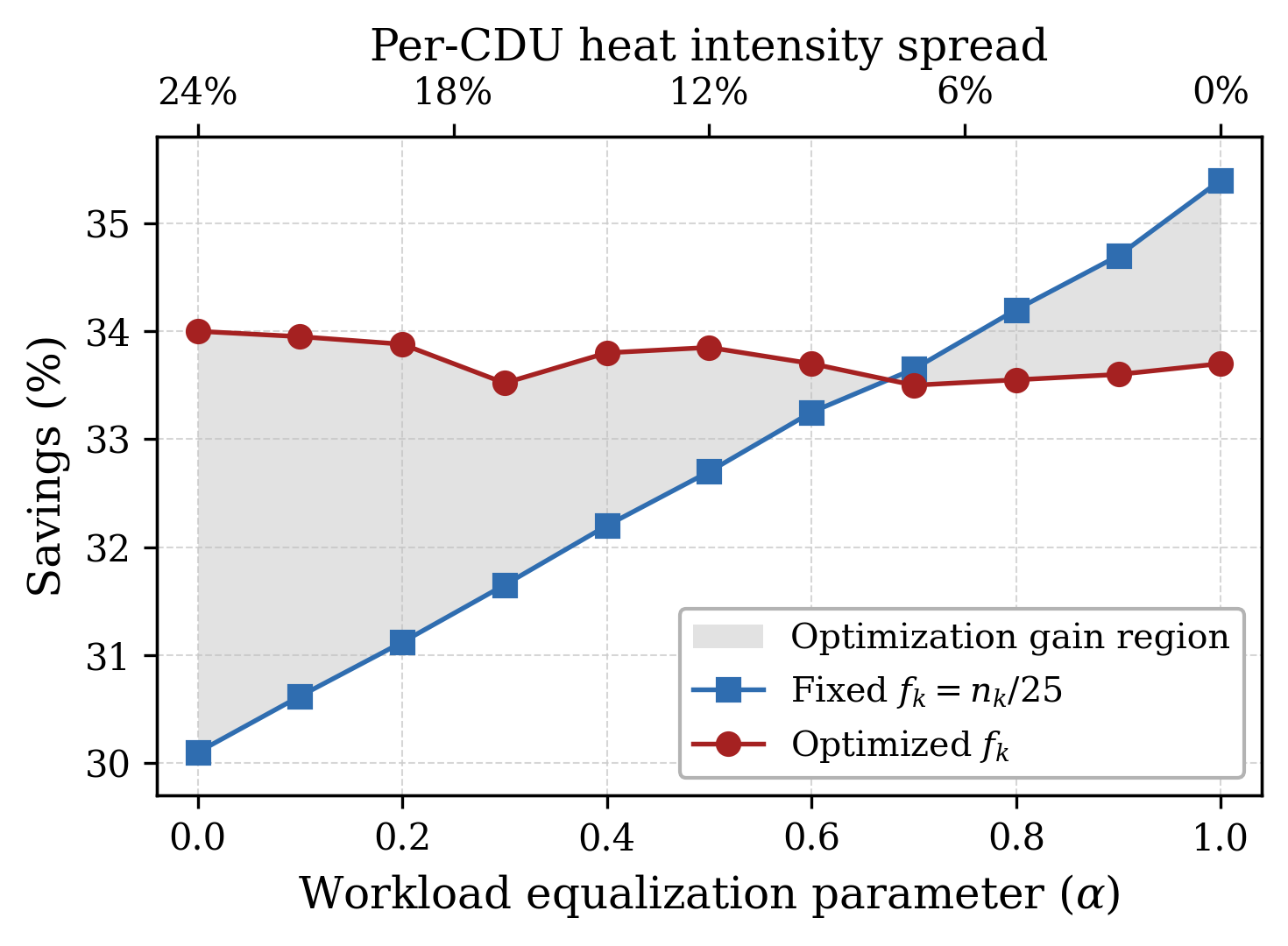}
\caption{Sensitivity of savings to the workload equalization parameter $\alpha$, for fixed and optimized flow fractions. The shaded region indicates the additional savings obtained from active flow optimization relative to fixed proportional flow.}
\label{fig:workload}
\end{figure}

The preceding results assume that the per-CDU computational heat load follows the empirical distribution measured at Frontier during 2023 \cite{Sun2024}, in which the heaviest CDU rejects approximately 24\% more heat per unit than the lightest CDU. Figure~\ref{fig:workload} examines how the optimization results change as this distribution is artificially equalized. The workload equalization parameter $\alpha \in [0, 1]$ interpolates between the empirical distribution at $\alpha = 0$ and a perfectly uniform distribution at $\alpha = 1$ in which every CDU rejects the same heat load. Formally, for each CDU $j \in \{1, \ldots, 25\}$ with empirical heat load $q_j^{\mathrm{emp}}$ and mean heat load $\bar{q} = (1/25) \sum_j q_j^{\mathrm{emp}}$, the modified heat load at equalization level $\alpha$ is
\begin{equation}
q_j(\alpha) = (1 - \alpha) \, q_j^{\mathrm{emp}} + \alpha \, \bar{q}
\end{equation}

The subloop heat loads used in the optimization are then obtained by summing $q_j(\alpha)$ over the CDUs assigned to each subloop, preserving mass balance for any partition size $K$. The per-CDU heat intensity spread, shown on the upper horizontal axis, decreases from 24\% at $\alpha=0$ to 0\% at $\alpha=1$.

Under fixed proportional flow fractions (blue curve), the savings increase monotonically with $\alpha$, rising from 30.1\% at $\alpha=0$ to 35.4\% at $\alpha=1$. This behavior is expected, because when all CDUs have identical heat loads, any partition and any proportional flow allocation produce the same per-subloop thermal intensity, and the design sensitivity to CDU assignment vanishes.

Under optimized flow fractions (red curve), the savings remain nearly constant across the full range of $\alpha$, varying by less than 1\%. This behavior demonstrates that the flow optimizer achieves through active per-timestep redistribution what workload equalization achieves passively through uniform thermal loading: both mechanisms equalize the per-subloop thermal intensity, and the energy benefit is bounded by the same physical limit. The shaded region between the two curves represents the additional savings obtained from active flow optimization relative to fixed proportional flow; this region is widest at $\alpha = 0$, where the imbalance is largest, and contracts to zero near $\alpha = 1$.

At high equalization levels ($\alpha \gtrsim 0.7$), the fixed-$f_k$ curve lies marginally above the optimized-$f_k$ curve. This crossover reflects the fact that, in the nearly uniform regime, the fixed proportional allocation $f_k = n_k / N$ is already very close to the thermal-intensity-equalizing allocation, and the SLSQP optimizer's finite convergence tolerance occasionally produces solutions below the fixed-$f_k$ solution. The crossover does not indicate that flow optimization produces less savings than proportional allocation in any physically meaningful sense; the two curves coincide to within the solver tolerance, and both correspond to the same underlying optimal operating point. The crossover is preserved in the figure rather than smoothed, to reflect the raw optimizer output without post-hoc adjustment.

\subsection{Sensitivity to Flow-Fraction Bounds}\label{sec:results:bounds}

The flow fraction bounds $0.05 \leq f_k \leq 0.95$ used in the main optimization are wider than the authority of typical hydraulic balancing hardware. To test whether the recommended designs depend on this range, Strategy~C was re-evaluated with the bounds tightened to $0.20 \leq f_k \leq 0.80$. The optimal partition $(19, 6)$ changes by less than 0.01\%, and the current Frontier partition $(14, 6, 5)$ changes by about 0.15\%. Both recommended designs hold their optimized flow fractions within roughly 3\% of the proportional values across the year, so they do not rely on extreme allocations and remain achievable within realistic actuator authority. Only highly imbalanced partitions, such as $(20, 3, 2)$, whose proportional split already places 8\% of the total flow in a single subloop, are materially affected by the bound width, and these partitions are not among the recommended designs. The feasibility of flow fraction optimization on the existing plant is therefore not contingent on access to the full nominal range.

\subsection{Per-Subloop Pump Model}\label{sec:results:pump}

The aggregate pump model of Eq.~\eqref{eq:pump_power} depends on the total flow rate and does not resolve the hydraulic resistance of individual subloop branches. To test the sensitivity of the results to this simplification, Strategy~C was re-evaluated with a per-subloop pump model in which each branch contributes power in proportion to its hydraulic resistance times the cube of its branch flow. The model reduces to the aggregate form at the proportional design point and adds a penalty for asymmetric allocation, with the penalty strength varied over a range that brackets the expected branch-resistance spread. Two results follow. First, because the penalty vanishes at the proportional split and the proportional split is always feasible, Strategy~C retains the Strategy~B feasible set, so the per-timestep recovery ratio remains at least one by construction under this pump model rather than by solver behavior; the small number of sub-unity cases reported in Section~\ref{sec:results:recovery} arise from the deployed actuator ramp limits and not from the aggregate pump simplification. Second, the 93\% reduction in design sensitivity is defined by the gap between the current $(14, 6, 5)$ and optimal $(19, 6)$ designs, both of which hold flow fractions within about 3\% of proportional; the per-subloop penalty therefore shifts the gap by only a few tenths of a MWh per year and does not change the conclusion. The pump term accounts for 27\% of the cooling energy at Frontier, with cooling tower fans accounting for the remaining 73\%, so a penalty acting on the pump term alone has limited influence over the total. The aggregate model is retained in the main results, and the per-subloop resistance is identified as a refinement in Section~\ref{sec:discuss:limitations}.

\subsection{Surrogate Model Uncertainty}\label{sec:results:uncertainty}

The optimal $(19, 6)$ design and the current $(14, 6, 5)$ design differ by 0.18\% in annual savings, which is small relative to the accuracy of the reduced-order surrogate. To test whether this margin is resolvable, the surrogate predictions were perturbed by the Stage~1 CV-RMSE of 2.31\%, applied as a first-order autoregressive process to preserve temporal correlation, and the two designs were re-evaluated over 3,000 Monte Carlo draws. The probability that the ranking of the two designs reverses is below 1\%. The savings of each design are moreover a ratio of two predictions from the same surrogate evaluated under the same loads, so any common-mode bias cancels and cannot move the ranking. The 0.18\% margin corresponds to approximately 3.3 MWh per year, which is both smaller than the surrogate absolute accuracy and economically negligible. The two designs are therefore effectively equivalent in energy terms, which reinforces rather than undermines the practical recommendation to retain the existing three-subloop topology.

\subsection{Recovery Ratio Analysis}\label{sec:results:recovery}

\begin{figure}[t]
\centering
\includegraphics[width=0.5\columnwidth]{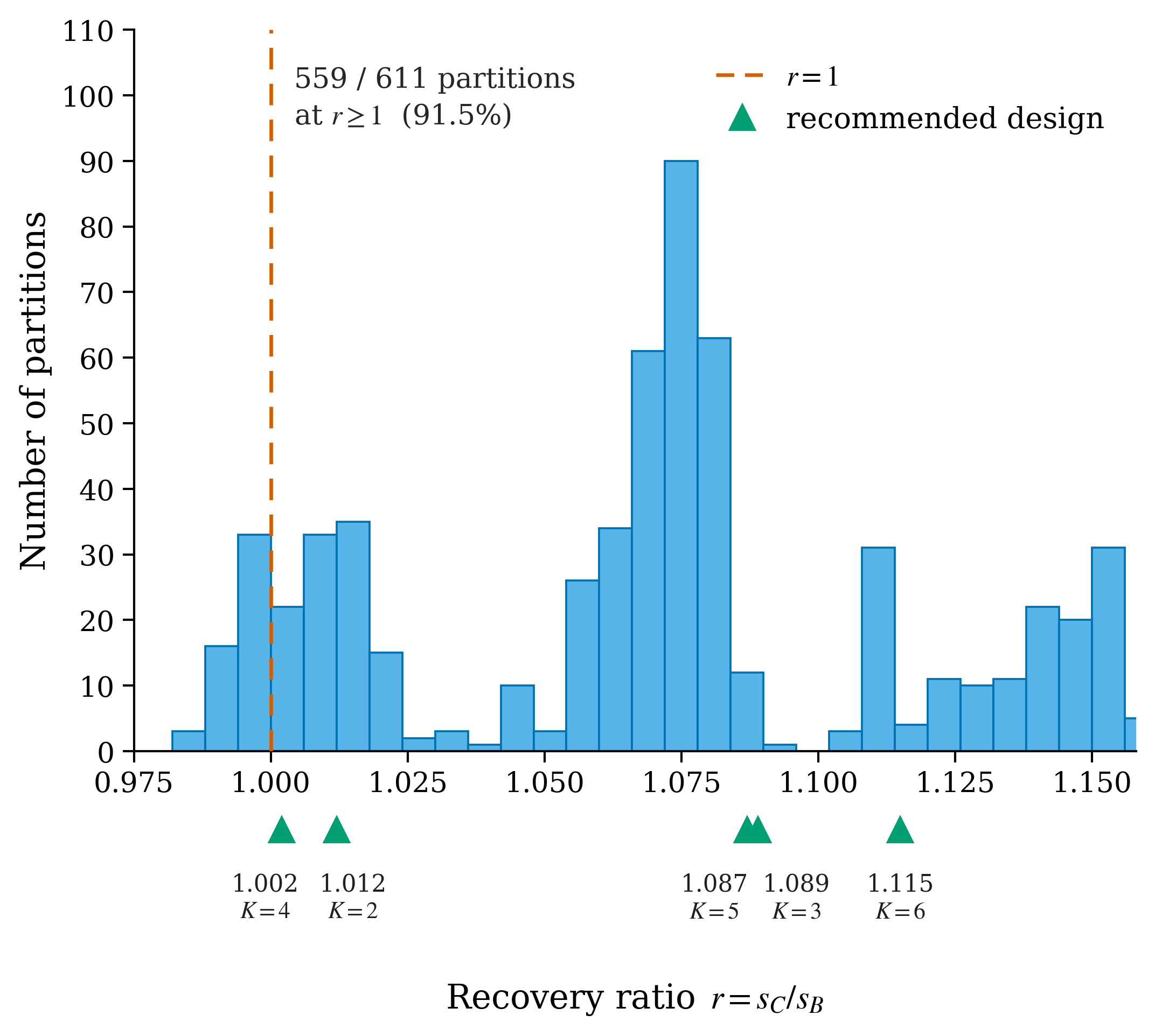}
\caption{Distribution of the recovery ratio $r = s_C / s_B$ across all 611 partitions under balanced CDU assignment, with Strategy~B taken as the fixed proportional allocation. The dashed line marks $r = 1$. The ratio exceeds unity for 91.5\% of partitions and for all recommended designs; the partitions below unity are near-balanced layouts where flow fraction optimization offers no benefit, and the deployed ramp limits carry a small penalty.}
\label{fig:recovery}
\end{figure}

Figure~\ref{fig:recovery} presents the distribution of the recovery ratio $r = s_C / s_B$ across all 611 partitions under balanced CDU assignment, where Strategy~B uses the fixed proportional flow fractions $f_k = n_k / N$ defined in Section~\ref{sec:method:strategies}. The ratio quantifies the fraction of the Strategy~B savings that Strategy~C is able to preserve once the flow fraction optimization is added as an additional degree of freedom. Under the per-timestep co-design formulation, $r \geq 1$ holds by construction, because Strategy~C extends the Strategy~B feasible set and can never produce a worse objective value than the proportional split it contains. Recomputed by this fixed proportional Strategy~B, the deployed Strategy~C satisfies $r \geq 1$ for 559 of the 611 partitions (91.5\%), including every best-in-class design in Table~\ref{tab:best_per_K}. The largest recovery occurs for the most thermally imbalanced partitions, where the proportional allocation is furthest from the thermal-intensity-equalizing allocation and flow fraction optimization is most valuable.

The mechanism by which Strategy~C outperforms Strategy~B at the majority of partitions is the following. Strategy~B constrains each subloop flow fraction to the proportional value $f_k = n_k / N$, which is thermally optimal only when the per-CDU heat load is uniform. When the per-CDU heat load varies across CDUs (as at Frontier, with a 24\% spread), the proportional allocation leaves some subloops with excess flow relative to their thermal load and others with insufficient flow. Strategy~C removes this constraint and permits the optimizer to equalize the per-subloop thermal intensity $w_k / f_k$ directly. This additional freedom reduces the binding thermal constraint, allows a lower total flow rate and a higher supply temperature setpoint to be selected at each timestep, and reduces the combined pump and cooling tower fan power.

The 52 partitions with $r < 1$ fall below unity by at most 1.7\% (minimum $r = 0.983$) and are all near-balanced layouts, such as $(7, 5, 5, 4, 4)$ and $(5, 5, 5, 5, 5)$, for which the proportional split already sits close to the thermal-intensity-equalizing allocation. For these layouts, the flow fraction freedom in Strategy~C yields no measurable benefit, while the actuator ramp limits carried by the deployed Strategy~C impose a small penalty that the ramp-free Strategy~B does not pay. The sub-unity values are therefore confined to the layouts where flow fraction optimization is not needed; for every partition with uneven subloop loads, and for all recommended designs, $r$ exceeds unity. An earlier version of this analysis compared Strategy~C through an intermediate Strategy~B variant that itself optimized the flow fractions, which understated the recovery; measured by the fixed proportional Strategy~B that the text defines, the imbalanced partitions that previously fell below unity all rise above it, with recovery ratios as high as 1.16.

From a deployment standpoint, the benefit of Strategy~C over the fixed proportional Strategy~B corresponds to approximately 10~MWh/yr of additional energy savings for the Frontier cooling plant. This benefit is modest compared to the 220~MWh/yr savings achieved by the underlying co-design optimization strategy, but it is obtained at essentially zero hardware cost provided the subloop pumps already support independent speed control, which is already the case for the Frontier HTW pumps. The principal value of flow fraction optimization is therefore not the additional annual energy it recovers but the design robustness it provides: it removes the dependence of plant performance on the CDU-to-subloop assignment quantified in Section~\ref{sec:results:flowopt}, and so makes the topology decision nearly immaterial to the energy outcome.

\section{Discussion}\label{sec:discussion}

\subsection{Baseline Characterization and Best-Practice Comparison}\label{sec:discuss:baseline}

A natural question is how much of the reported 35\% saving reflects genuine optimization rather than a conservative baseline. The measured baseline at Frontier runs a mean supply temperature of 20.0$^{\circ}$C, with a 5th to 95th percentile range of 11.9 to 28.1$^{\circ}$C, and per-subloop return temperatures averaging 28 to 34$^{\circ}$C, leaving 8 to 14$^{\circ}$C of mean margin to the 42$^{\circ}$C limit. The mean supply temperature sits 22$^{\circ}$C below the limit, and the total flow is close to twice the value required to hold the binding return temperature, so the baseline carries substantial headroom in both actuators. The baseline is not an idealized supervisory controller: measured return temperatures occasionally exceed the 42$^{\circ}$C limit, reaching 65.9$^{\circ}$C in one subloop during a high-load period, whereas all of the optimized strategies respect the limit by construction and are therefore safer as well as more efficient.

To separate the contribution of the proposed framework from that of standard best practice, two reference controllers were evaluated. A supply-temperature-reset controller that raises the supply temperature by the thermal limit while holding the measured flow recovers 15.7\% of the baseline cooling energy. Flow reduction alone, which is Strategy~A, recovers 23.4\%. The combined fixed-fraction co-design, which is Strategy~B, recovers 34.8\%, and the full three-layer Strategy~C recovers 35.30\% at the current topology. The framework therefore delivers about 12\% beyond the strongest single best-practice lever. A large part of the headline saving does reflect baseline over-pumping and a conservative supply temperature, which conventional supervisory control would also capture; the distinct contribution of the three-layer framework is the flow fraction layer, whose value is the design robustness documented in Section~\ref{sec:results:flowopt} rather than a large increment of raw energy. Stating this explicitly clarifies what the framework adds over conventional practice and what it does not.

\subsection{Practical Implications for Plant Design}\label{sec:discuss:practical}

The results establish a clear hierarchy of design decisions for liquid-cooled data center cooling plants. The most consequential decision is the operational strategy: moving from flow-only optimization (Strategy~A, 23.4\% savings) to joint flow and temperature co-design optimization with optimized flow fractions (Strategy~C, 35.48\% savings at the globally optimal partition) yields a 12.0\% improvement, equivalent to approximately 220~MWh/yr for the Frontier cooling plant. In contrast, the choice of subloop count $K$ affects savings by at most 0.18\% between the best $K=2$ partition (35.48\%) and the best $K=3$ partition (35.44\%, which includes the current Frontier design at 35.30\%). The CDU-to-subloop allocation within a given $K$ affects savings by at most 0.2\% when flow fractions are optimized.

This hierarchy has a direct practical consequence. The current Frontier topology is a $K=3$ design with partition $(14, 6, 5)$, which achieves 35.30\% savings under Strategy~C. The theoretical global optimum is the $K=2$ partition $(19, 6)$ at 35.48\%, a difference of 0.18\% or approximately 3.3~MWh/yr. Converting from three subloops to two would require extensive piping modifications, decommissioning of a subloop pump and heat exchanger, and recommissioning of the thermal safety interlocks. In contrast, enabling flow fraction optimization on the existing three-subloop topology produces approximately 10~MWh/yr of additional savings relative to the proportional allocation baseline without any hardware modification beyond a control system update. The software-only retrofit is therefore both less disruptive and more cost-effective than the topology-level redesign.

\subsection{Simultaneous Optimization of Flow Fractions and Workload}\label{sec:discuss:substitutability}

The present analysis treats flow fraction optimization (which we refer to as Layer~2 in the three-layer framework) and workload scheduling across subloops (Layer~3) as separate optimization layers. The sensitivity analysis in Section~\ref{sec:results:workload} shows that when flow fractions are optimized, the total energy savings are nearly invariant to the workload distribution, and when workload is equalized, the total energy savings are nearly invariant to whether the flow fractions are proportional or SLSQP-optimized. These observations suggest that the two layers address the same underlying physical degree of freedom, namely the equalization of the per-subloop thermal intensity $w_k / f_k$.

A more complete formulation would optimize the flow fractions and the workload allocation simultaneously in a single optimization problem, treating both $f_k$ and the CDU-to-subloop heat assignment $w_k$ as decision variables subject to the total flow mass balance $\sum_k f_k = 1$ and the total heat load mass balance $\sum_k w_k = Q_{\mathrm{tot}}$. Under the ideal assumption that workload can be freely redistributed across subloops at every timestep, the simultaneous optimization would produce the same energy minimum as either layer alone, because both layers can independently achieve the thermal intensity equalization condition. In practice, however, workload redistribution is constrained by application locality requirements, interconnect topology (the Slingshot dragonfly network in Frontier), and job scheduler overhead, whereas flow fraction adjustment requires only variable-speed pumps or modulating valves on each subloop branch. The simultaneous formulation would allow the optimizer to exploit whichever layer is feasible at each timestep, using workload redistribution when the scheduler permits it and flow redistribution otherwise. This hybrid formulation is left for future work and is discussed further in Section~\ref{sec:discuss:limitations}.

\subsection{Role of CDU Assignment Under Optimized Flow}\label{sec:discuss:assignment}

The near-elimination of the CDU assignment effect under optimized flow fractions (Fig.~\ref{fig:assignment_gap}) has implications for cooling plant commissioning. In current practice, CDU-to-subloop assignment is typically determined during installation based on physical proximity and piping convenience, with limited attention to thermal balancing. The results show that this approach incurs an energy penalty of up to 5.3\% under fixed flow fractions because it creates subloops with systematically different heat intensities. Flow fraction optimization largely neutralizes this penalty, relaxing the commissioning requirement from thermal balancing to ensuring that each subloop branch has independently controllable flow.

The residual 0.4\% design sensitivity under optimized $f_k$ is not zero. The remaining variability arises from nonlinear interactions among the number of CDUs per subloop ($n_k$), the per-CDU heat load, and the heat exchanger effectiveness. Even with optimal flow distribution, subloops with very few CDUs (for example $n_k = 1$ or $n_k = 2$) operate at different effectiveness values than subloops with many CDUs, introducing a second-order performance difference that flow optimization cannot fully eliminate.

\subsection{Recovery Ratio Interpretation}\label{sec:discuss:recovery}

The recovery ratio $r = s_C / s_B$ exceeds unity for 91.5\% of the 611 partitions under balanced CDU assignment, and for all recommended designs, when Strategy~B is taken as the fixed proportional allocation defined in Section~\ref{sec:method:strategies}. Under the per-timestep co-design formulation, $r \geq 1$ is expected, because Strategy~C extends the Strategy~B feasible set by removing the proportional flow constraint, and the SLSQP solution of the extended problem can therefore never produce a worse objective value than the solution of the restricted problem. The improvement reflects the cumulative benefit over 49{,}353 timesteps of allowing the optimizer to adjust flow fractions dynamically in response to the instantaneous per-CDU heat load distribution, rather than being locked into a single fixed allocation for the full year.

The partitions with $r$ marginally below unity, reported as 5.2\% compared to a flow-optimized Strategy~B variant in an earlier version, are near-balanced layouts where the proportional split is already near optimal; there the deployed Strategy~C pays a small ramp-limit penalty while gaining nothing from flow fraction freedom, as detailed in Section~\ref{sec:results:recovery}. For the Frontier cooling plant, the deployment-relevant conclusion is that Strategy~C provides approximately 10~MWh/yr of additional savings over the fixed proportional Strategy~B at essentially zero hardware cost, and this benefit is positive at every recommended design and at every partition with uneven subloop loads.

\subsection{Limitations and Future Work}\label{sec:discuss:limitations}

Several limitations of the present study merit discussion. First, the physics model uses a reduced-order representation of the cooling tower, pump, and heat exchanger subsystems, calibrated through the Stage~1 Modelica digital twin rather than directly through the physical plant. While the Stage~1 model was validated through one full year of operational data following ASHRAE Guideline~14 \cite{ASHRAE2014} (CV-RMSE between 1.96\% and 2.67\%, normalized mean bias error within $\pm 2.5\%$), the Stage~2 framework inherits any systematic biases present in that calibration. The reduced-order model is itself validated through the Stage~1 digital twin, both at the calibration point and under the optimized off-baseline schedule, in Appendix~A.

Second, the analysis assumes that per-CDU heat loads are exogenous and follow the empirical distribution observed in 2023 \cite{Sun2024}. HPC workload patterns may shift as new applications are deployed on Frontier, potentially altering the optimal partition and flow allocation. The workload equalization sensitivity analysis in Section~\ref{sec:results:workload} provides bounds on performance across a range of distributions, but a fully stochastic treatment of workload uncertainty is left for future work.

Third, the flow fraction optimization assumes that per-subloop flow control is technically feasible with the existing piping infrastructure. The Frontier cooling plant uses four independent variable-speed pumps (three active and one standby) serving three subloops, with empirical flow fractions of approximately 24.6\%, 26.0\%, and 49.5\%. These fractions are not proportional to the CDU counts in the current partition $(14, 6, 5)$, indicating that some degree of independent flow control already exists at the hardware level. The full range of flow fractions assumed by the optimizer ($0.05 \leq f_k \leq 0.95$) may not be achievable at every operating point without control system modifications, and the effective range of feasible flow fractions in deployment may be narrower than the range assumed in the offline optimization. The bound-sensitivity analysis in Section~\ref{sec:results:bounds} addresses this directly by tightening the bounds to $0.20 \leq f_k \leq 0.80$, which leaves the recommended designs essentially unchanged, because they hold their flow fractions within about three\% of proportional, so the conclusions do not depend on access to the full nominal range.

Future work will address four extensions. First, closed-loop validation of the flow fraction optimization through Modelica co-simulation, in which the digital twin runs in closed loop with the SLSQP optimizer, extending the open-loop off-baseline validation of Appendix~A to verify that the reduced-order performance predictions are consistent with the high-fidelity physics model under feedback. Second, simultaneous optimization of Layer~2 (flow fractions) and Layer~3 (workload distribution) in a single formulation, as discussed in Section~\ref{sec:discuss:substitutability}, to quantify the additional benefit of combining the two mechanisms under realistic scheduler constraints. Third, extension to time-varying partition assignment, in which the CDU-to-subloop mapping can change seasonally to track shifting workload patterns. Fourth, generalization to other liquid-cooled HPC facilities to assess whether the $K=2$ optimality result and the flow-workload equivalence finding hold across different plant architectures and scales.

\section{Conclusion}\label{sec:conclusion}

This paper addressed the cooling plant design problem for the Frontier exascale supercomputer by extending the Stage~1 digital twin~\cite{Jadhav2026Stage1} from a fixed-topology control study to a joint topology and operational optimization. The plant design space was structured as a three-layer problem: the integer partition of 25 CDUs across $K$ parallel subloops (Layer~1), the continuous flow fraction allocation across subloops (Layer~2), and the per-timestep co-design optimization of total flow rate and supply temperature setpoint (Layer~3). The 611 feasible partitions for $K$ between 2 and 6 were enumerated exhaustively and paired with SLSQP on the continuous subproblem, providing a global optimality certificate over the discrete design space.

Four principal findings emerge from the analysis. First, the optimal plant design within the enumerated surrogate problem is a two-subloop configuration with partition $(19, 6)$, achieving 35.48\% annual cooling energy savings relative to the measured baseline. The current three-subloop Frontier design with partition $(14, 6, 5)$ achieves 35.30\% savings under the same operational strategy, placing it 0.18\% below the global optimum. Second, the energy benefit within the optimal $K=2$ configuration is insensitive to the CDU-to-subloop allocation when flow fractions are optimized: the 12 feasible $K=2$ partitions span a range of only 0.20\%, and the CDU assignment gap between balanced and worst-case ordering collapses from 2 to 3\% under fixed proportional flow to approximately 0.1\% under optimized flow. Third, flow fraction optimization acts as a functional equivalent of workload equalization: the two mechanisms equalize the per-subloop thermal intensity through different physical pathways and converge to the same energy minimum, with flow optimization being the more practical of the two because it requires only a control system update rather than coordination with the HPC job scheduler. Fourth, the design decision hierarchy that emerges from the results places the choice of operational strategy well above the topology decisions in energy impact: the gap between flow-only optimization and full three-layer co-design optimization is approximately 12\%, while the gap between the best and worst topology decisions within the optimized-flow regime is less than 0.5\%.

For the Frontier facility, these findings translate directly into a deployment recommendation. The existing three-subloop topology should be retained rather than reconfigured, because the 3.3~MWh/yr energy advantage of a two-subloop redesign is small relative to the capital and disruption cost of decommissioning a subloop pump and heat exchanger. The primary near-term improvement available to the facility is activating independent per-subloop flow control, which recovers approximately 10~MWh/yr relative to the fixed proportional baseline and requires no hardware modification beyond a supervisory control update. This recommendation is specific to Frontier, but the underlying framework is transferable. For any liquid-cooled HPC plant with multiple parallel subloops and independently controllable pumps, the three-layer optimization can be applied with site-specific recalibration to identify the equivalent deployment priority.

Beyond the immediate application to Frontier, the present work contributes a methodological template for cooling plant design optimization in which the integer topology variables are small enough to enumerate exhaustively, and the continuous operational variables admit analytical gradients. This structure is not unique to HPC facilities; it arises in chiller plants with multiple parallel chillers, in district cooling networks with multiple production nodes, and in heat exchanger networks with a small number of candidate stream matches. The decomposition approach adopted here, namely complete enumeration over the integer variables combined with SLSQP on the continuous subproblem, offers a provably optimal and computationally tractable alternative to general mixed-integer nonlinear programming for problems in this class, and the global optimality certificate obtained through enumeration is a property that branch-and-bound and outer-approximation methods cannot guarantee on nonconvex thermal-hydraulic models.

Together with the Stage~1 digital twin paper~\cite{Jadhav2026Stage1} and the earlier data-driven inefficiency diagnosis~\cite{Jadhav2026ML}, the present work completes a three-stage progression from empirical waste quantification, to validated physics-based control optimization, to design and operational co-design optimization. The three stages share a single operational dataset~\cite{Sun2024} and a consistent set of calibrated parameters, and together they document the full range of energy improvements available to the Frontier cooling plant without hardware replacement: from the 85~MWh of waste identified by the machine learning surrogate, through the 20 to 30\% savings achievable by control-layer optimization, to the 35.48\% upper bound established in the present plant-level analysis. As exascale computing facilities continue to scale toward multi-facility deployments and multi-gigawatt power envelopes, the integration of data-driven diagnosis, validated digital twins, and layered optimization frameworks provides a replicable pathway for closing the gap between current operation and the thermodynamic limits of liquid-cooled heat rejection.

\section*{Acknowledgments}
This research was partially supported by the University of Michigan-Dearborn Office of Research through Research Initiation \& Development (RID) Grant.

\section*{Data and Code Availability}
The data and code supporting the findings of this study are publicly available on GitHub at \url{https://github.com/m-iml/Co-Design_Optimization_Data_Center_Digital_Twin}.

\section*{Appendix A: Validation of the Reduced-Order Model}

The reduced-order model used throughout this paper was validated through the Stage~1 Modelica digital twin built on the Buildings Library, which uses the temperature-dependent Melinder property correlations for the ethylene glycol-water coolant. The validation was carried out in two parts.
First, the per-subloop return temperature predictions of the reduced-order energy balance were compared with the digital twin across six candidate designs, spanning $K = 2$ and $K = 3$ topologies, over the full 2023 operational record. The reduced-order predictions agree with the twin with a systematic bias of $-0.36^{\circ}$C, a root mean squared error of $0.40^{\circ}$C, and a coefficient of determination of $R^2 = 0.9916$. The bias is constant across all six designs and is fully explained by the difference between the constant specific heat of 3,500~J/(kg$\cdot$K) used in the reduced-order model and the temperature-dependent value in the Buildings Library correlation. Because the bias is constant across designs, it does not affect their relative ranking.
Second, to confirm that the model remains valid under the optimized off-baseline operating regime on which the conclusions depend, the digital twin was driven by the Strategy~C optimized schedule for the current and optimal designs, with the supply temperature near 28$^{\circ}$C and the total flow reduced to about 188~kg/s, well away from the measured baseline of 20$^{\circ}$C and 257~kg/s. Under this optimized schedule, the twin confirms that every subloop return temperature remains at or below the 42$^{\circ}$C thermal limit, so the constraint satisfaction reported by the reduced-order optimizer is preserved under the full thermodynamic property model. The reduced-order predictions are therefore trustworthy both at the calibration point and across the operating envelope explored by the optimization.

%%%%%%%%%%%%%  BIBLIOGRAPHY  %%%%%%%%%%%%%%%%%%%%%%%%%%%%%%%%%%%%%%%%%

\bibliographystyle{unsrtnat}

\bibliography{references_verified} %% verified bibliography

@article{Sun2024,
  author  = {Sun, Jian and Gao, Zhiming and Grant, David and Nawaz, Kashif and Wang, Pengtao and Yang, Cheng-Min and Boudreaux, Philip and Kowalski, Stephen and Huff, Shean},
  title   = {Energy dataset of {Frontier} supercomputer for waste heat recovery},
  journal = {Scientific Data},
  volume  = {11},
  number  = {1},
  pages   = {1077},
  year    = {2024},
  doi     = {10.1038/s41597-024-03913-w}
}

@techreport{ASHRAE2014,
  author      = {{ASHRAE}},
  title       = {{ASHRAE} {Guideline} 14-2014: Measurement of Energy, Demand, and Water Savings},
  institution = {American Society of Heating, Refrigerating and Air-Conditioning Engineers},
  address     = {Atlanta, GA},
  year        = {2014}
}

@article{Jadhav2026Stage1,
  author        = {Jadhav, Shrenik and Liu, Zheng},
  title         = {Digital Twin-Based Cooling System Optimization for Data Center},
  journal       = {arXiv preprint},
  year          = {2026},
  month         = mar,
  eprint        = {2603.01198},
  archive       = {arXiv},
  archivePrefix = {arXiv},
  primaryClass  = {eess.SY},
  doi           = {10.48550/arXiv.2603.01198}
}

@article{Jadhav2026ML,
  author        = {Jadhav, Shrenik and Liu, Zheng},
  title         = {Machine Learning Guided Cooling System Optimization for Data Center},
  journal       = {arXiv preprint},
  year          = {2026},
  month         = jan,
  eprint        = {2601.02275},
  archive       = {arXiv},
  archivePrefix = {arXiv},
  primaryClass  = {eess.SY},
  doi           = {10.48550/arXiv.2601.02275}
}

@article{Virtanen2020,
  author  = {Virtanen, Pauli and Gommers, Ralf and Oliphant, Travis E. and Haberland, Matt and Reddy, Tyler and Cournapeau, David and Burovski, Evgeni and Peterson, Pearu and Weckesser, Warren and Bright, Jonathan and {van der Walt}, St{\'e}fan J. and others},
  title   = {{SciPy} 1.0: Fundamental Algorithms for Scientific Computing in {Python}},
  journal = {Nature Methods},
  volume  = {17},
  pages   = {261--272},
  year    = {2020},
  doi     = {10.1038/s41592-019-0686-2}
}

@inproceedings{Karimi2024,
  author    = {Karimi, Ahmad Maroof and Maiterth, Matthias and Shin, Woong and others},
  title     = {Exploring the frontiers of energy efficiency using power management at system scale},
  booktitle = {SC '24 Workshops of the International Conference for High Performance Computing, Networking, Storage and Analysis},
  pages     = {1835--1844},
  year      = {2024},
  doi       = {10.1109/SCW63240.2024.00230}
}

@inproceedings{Grant2026,
  author    = {Grant, Dustin and Bortot, Lorenzo and DePrater, Clint and Martinez, David and Grant, Ryan and Bates, Natalie},
  title     = {Providing thermal stability for an exascale supercomputer: A case study of {Frontier}'s cooling system},
  booktitle = {Proceedings of Supercomputing Asia and ICHPC},
  pages     = {69--78},
  year      = {2026},
  doi       = {10.1145/3784828.3785159}
}

@article{Wang2024P,
  author  = {Wang, Peng and Kowalski, Steven and Gao, Zhiming and others},
  title   = {District heating utilizing waste heat of a data center: High-temperature heat pumps},
  journal = {Energy and Buildings},
  volume  = {315},
  pages   = {114327},
  year    = {2024},
  doi     = {10.1016/j.enbuild.2024.114327}
}

@inproceedings{Brewer2024,
  author    = {Brewer, Wesley and Maiterth, Matthias and Kumar, Vivek and others},
  title     = {A digital twin framework for liquid-cooled supercomputers as demonstrated at exascale},
  booktitle = {SC '24: Proceedings of the International Conference for High Performance Computing, Networking, Storage and Analysis},
  articleno = {23},
  pages     = {1--18},
  year      = {2024},
  doi       = {10.1109/SC41406.2024.00029}
}

@inproceedings{Kumar2024,
  author    = {Kumar, Vivek and Greenwood, Scott and Brewer, Wesley and Grant, Dustin and Parkison, Nicholas and Williams, Wayne},
  title     = {Thermo-fluid modeling framework for supercomputing digital twins: {Part 1}, demonstration at exascale},
  booktitle = {Proceedings of the American Modelica Conference},
  year      = {2024}
}

@inproceedings{Greenwood2024,
  author    = {Greenwood, Scott and Kumar, Vivek and Brewer, Wesley},
  title     = {Thermo-fluid modeling framework for supercomputing digital twins: {Part 2}, automated cooling models},
  booktitle = {Proceedings of the American Modelica Conference},
  year      = {2024}
}

@techreport{IEA2025,
  author      = {{International Energy Agency}},
  title       = {Energy and {AI}},
  institution = {IEA},
  address     = {Paris},
  year        = {2025},
  note        = {IEA Special Report}
}

@techreport{Shehabi2024,
  author      = {Shehabi, Arman and Smith, Sarah J. and Hubbard, Amanda and others},
  title       = {2024 {United States} Data Center Energy Usage Report},
  institution = {Lawrence Berkeley National Laboratory},
  number      = {LBNL-2001637},
  year        = {2024},
  doi         = {10.71468/P1WC7Q}
}

@article{Zhang2021,
  author  = {Zhang, Qinghui and Tang, Chao and Bai, Ting and others},
  title   = {A survey on data center cooling systems: Technology, power consumption modeling and control strategy optimization},
  journal = {Journal of Systems Architecture},
  volume  = {119},
  pages   = {102253},
  year    = {2021},
  doi     = {10.1016/j.sysarc.2021.102253}
}

@article{Ebrahimi2014,
  author  = {Ebrahimi, Khosrow and Jones, Gerard F. and Fleischer, Amy S.},
  title   = {A review of data center cooling technology, operating conditions and the corresponding low-grade waste heat recovery opportunities},
  journal = {Renewable and Sustainable Energy Reviews},
  volume  = {31},
  pages   = {622--638},
  year    = {2014},
  doi     = {10.1016/j.rser.2013.12.007}
}

@techreport{UptimeInstitute2024,
  author      = {{Uptime Institute}},
  title       = {Global Data Center Survey Results 2024},
  institution = {Uptime Institute},
  year        = {2024}
}

@article{Wetter2014,
  author  = {Wetter, Michael and Zuo, Wangda and Nouidui, Thierry S. and Pang, Xiufeng},
  title   = {{Modelica Buildings} library},
  journal = {Journal of Building Performance Simulation},
  volume  = {7},
  number  = {4},
  pages   = {253--270},
  year    = {2014},
  doi     = {10.1080/19401493.2013.765506}
}

@article{Fu2019a,
  author  = {Fu, Yangyang and Zuo, Wangda and Wetter, Michael and VanGilder, James W. and Han, Xu and Plamondon, David},
  title   = {Equation-based object-oriented modeling and simulation for data center cooling: A case study},
  journal = {Energy and Buildings},
  volume  = {186},
  pages   = {108--125},
  year    = {2019},
  doi     = {10.1016/j.enbuild.2019.01.017}
}

@article{Fu2019b,
  author  = {Fu, Yangyang and Zuo, Wangda and Wetter, Michael and VanGilder, James W. and Yang, Peilin},
  title   = {Equation-based object-oriented modeling and simulation of data center cooling systems},
  journal = {Energy and Buildings},
  volume  = {198},
  pages   = {503--519},
  year    = {2019},
  doi     = {10.1016/j.enbuild.2019.06.037}
}

@article{Hinkelman2022,
  author  = {Hinkelman, Kathryn and Wang, Jing and Zuo, Wangda and others},
  title   = {Modelica-based modeling and simulation of district cooling systems: A case study},
  journal = {Applied Energy},
  volume  = {311},
  pages   = {118654},
  year    = {2022},
  doi     = {10.1016/j.apenergy.2022.118654}
}

@article{Fan2021,
  author  = {Fan, Cong and Hinkelman, Kathryn and Fu, Yangyang and others},
  title   = {Open-source {Modelica} models for the control performance simulation of chiller plants with water-side economizer},
  journal = {Applied Energy},
  volume  = {299},
  pages   = {117337},
  year    = {2021},
  doi     = {10.1016/j.apenergy.2021.117337}
}

@article{Grahovac2023,
  author  = {Grahovac, Milica and Ehrlich, Paul and Hu, Jianjun and Wetter, Michael},
  title   = {Model-based data center cooling controls comparative co-design},
  journal = {Science and Technology for the Built Environment},
  volume  = {30},
  number  = {4},
  pages   = {394--414},
  year    = {2023},
  doi     = {10.1080/23744731.2023.2276011}
}

@article{Lu2019,
  author  = {Lu, Yujie and Wang, Shengwei and Sun, Yongjun and Yan, Chengchu},
  title   = {Optimal Scheduling of Buildings with Energy Generation and Thermal Energy Storage Under Dynamic Electricity Pricing Using Mixed-Integer Nonlinear Programming},
  journal = {Applied Energy},
  year    = {2019}
}

@article{Huang2020,
  author  = {Huang, Sen and Zuo, Wangda and Sohn, Michael D.},
  title   = {A {Bayesian} Network Model for the Optimization of a Chiller Plant's Condenser Water Loop},
  journal = {Journal of Building Performance Simulation},
  year    = {2020}
}

@article{Afroz2018,
  author  = {Afroz, Zakia and Shafiullah, G. M. and Urmee, Tania and Higgins, Gary},
  title   = {Modeling Techniques Used in Building {HVAC} Control Systems: A Review},
  journal = {Renewable and Sustainable Energy Reviews},
  volume  = {83},
  pages   = {64--84},
  year    = {2018},
  doi     = {10.1016/j.rser.2017.10.044}
}

@article{Ma2012,
  author  = {Ma, Zhenjun and Wang, Shengwei},
  title   = {Supervisory and Optimal Control of Central Chiller Plants Using Simplified Adaptive Models and Genetic Algorithm},
  journal = {Applied Energy},
  volume  = {88},
  number  = {1},
  pages   = {198--211},
  year    = {2012},
  doi     = {10.1016/j.apenergy.2010.07.036}
}

@article{Furman2002,
  author  = {Furman, Kevin C. and Sahinidis, Nikolaos V.},
  title   = {A Critical Review and Annotated Bibliography for Heat Exchanger Network Synthesis in the 20th Century},
  journal = {Industrial and Engineering Chemistry Research},
  volume  = {41},
  number  = {10},
  pages   = {2335--2370},
  year    = {2002},
  doi     = {10.1021/ie010389e}
}

@article{Escobar2013,
  author  = {Escobar, Marcelo and Trierweiler, Jorge O.},
  title   = {Optimal Heat Exchanger Network Synthesis: A Case Study Comparison},
  journal = {Applied Thermal Engineering},
  volume  = {51},
  number  = {1-2},
  pages   = {801--826},
  year    = {2013},
  doi     = {10.1016/j.applthermaleng.2012.10.022}
}

%%%%%%%%%%%%%%%%%%%%%%%%%%%%%%%%%%%%%%%%%%%%%%%%%%%%%%%%%%%%%%%%%%%%%%

\end{document}